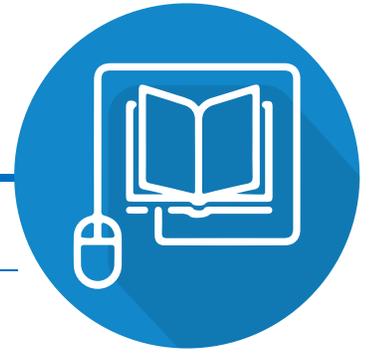

**Unite Paper 2021** <<number to be assigned>>

# A Framework for Ethical AI at the United Nations

| | |
|---|---|
| Prepared by: | **Lambert Hogenhout** |
| Organization: | **UN Office for Information and Communications Technology** |
| Contact: | **hogenhout@un.org** |
| Date: | **15-3-2021** |

### Contents

# Executive Summary

This paper aims to provide an overview of the ethical concerns in artificial intelligence (AI) and the framework that is needed to mitigate those risks, and to suggest a practical path to ensure the development and use of AI at the United Nations (UN) aligns with our ethical values.  The overview discusses how AI is an increasingly powerful tool with potential for good, albeit one with a high risk of negative side-effects that go against fundamental human rights and UN values. It explains the need for ethical principles for AI aligned with principles for data governance, as data and AI are tightly interwoven. It explores different ethical frameworks that exist and tools such as assessment lists. It recommends that the UN develop a framework consisting of ethical principles, architectural standards, assessment methods, tools and methodologies, and a policy to govern the implementation and adherence to this framework, accompanied by an education program for staff.

# Introduction

Artificial intelligence (AI) has become ubiquitous in our lives—from advertisements that target us as we browse the web, to autopilot features in cars, airplanes and public transport, to algorithms that screen job applications; and almost every week, the press informs us with excitement about new applications or achievements of AI).

The relentless digitization of the world also plays a role. Prompted by remote learning during the pandemic, a school in Hong Kong now uses an AI to read children's emotions from their video image as they learn[1], in order to assess how well they understood the material. And AIs are able to handle increasingly sophisticated tasks. Autonomous driving in real-life traffic, for instance, requires a dazzling range of interpretation and decision making.

Given that AI-based systems have access to and can process infinitely more information than any human, they are bound to eventually take better decisions than humans. In a number of fields this is already undeniably true: For example, there are AIs for various forms of cancer detection that outperform the average trained specialist. That list will continue to grow, and we will increasingly hand over decisions to AI. The AI being "better" is often defined as faster, more accurate, or more optimal according to certain criteria. But is that all that matters?

The immediate factors that the AI is programmed to achieve or optimize (example: "get the car from point A to point B") cannot not be the only considerations. We can impose constraints ("… and don't kill any pedestrians on the way") but they address only specific problems that may arise. As AIs get increasingly complex and handle a variety of unpredictable situations autonomously, the only way to ensure the AI does not do anything "bad", is to outfit them with a set of general morals and values. We need to build an ethical framework into our AI.

Stories about advanced AIs creating havoc, even trying to eliminate mankind, have been the topics of science fiction novels for a long time. While the "AI takes control" scenarios are unlikely for the foreseeable future, some concerns are quickly becoming reality as AI is increasingly used in financial systems, law enforcement systems, autonomous cars and weapons.

The UN Secretariat has already started using AI in some forms - Alba, a chatbot available to UN personnel, is one form of AI. These applications may not seem to be complex enough to need their own ethical framework, but as we will see further in this paper, some potentially problematic issues are already manifesting themselves; and in the coming years that need for an ethics framework for AI will become essential.

And indeed, many governments, regional organizations and businesses have started to consider the way they are using AI. Many have stated principles, and some have issued policies. The European Commission, for instance, is one of the main global players in the area of AI policymaking.

---

[1] CNN, 16 Feb 2021. See https://edition.cnn.com/2021/02/16/tech/emotion-recognition-ai-education-spc-intl-hnk/index.html

From a corporate perspective - other factors may play a role: the desire to be a good corporate citizen (which, as has been demonstrated, makes a lot of commercial sense too), audibility (what are we doing, based on which decisions), fiduciary responsibility or compliance.

Apart from preventing undesirable effects of AI, an ethics framework could also outline the positive contributions an AI should aim to make. From a UN perspective, the potential of AI for Good has been studied and debated (and implemented) for a number of years[2] and it deserves careful consideration to incorporate in a set of ethical principles.

The aim of this paper is to outline the ethical concerns of AI, now and in the near future, what the UN can and should do and how to go about it.

This paper is composed of three parts:

> **Part1**: An analysis of the ethical problems with AI: Why should we be concerned? What are the issues?
>
> **Part 2**: Defining Ethical AI: A detailed analysis of the ethics of AI, including a stock-take of what has been done
>
> **Part 3**: Implementing AI: Proposing a practical way forward for the UN. Includes an overview of research, examples and current tools

The intention of this paper is that it will lead to discussion followed by action, whether in the form of a set of guidelines, policy, a code of ethics, or educational initiatives to create awareness of the issues. This affects not only the technologists who may develop AI-based systems. It also concerns those involved in procuring or integrating AI systems, those at the managerial level who approve such projects, and ultimately the users affected by the systems. All these stakeholders need to be aware of the issues and the measures the UN intends to take, and all need to be taken into account when developing the ethics framework that guides our development and use of AI.

---

[2]   The AI for Good Global Summit organized yearly by ITU is one of the platforms where this is explored. see: https://aiforgood.itu.int

# 1. Problems with AI

## A. What are the risks of AI?

When an AI system gives a wrong medical diagnosis or the facial recognition in your smartphone fails, the concern is clear: the system does not accurately perform its task, it makes a mistake. We can of course wonder to what extent this risk is acceptable. After all, humans also make mistakes. The captain of the Titanic sailed too close to an iceberg with disastrous results. And at more personal level, you may have once forgotten your umbrella in the subway or left your dinner on the stovetop for too long, causing it to burn. We accept some of those mistakes as an inevitable fact of life, so to what extent should we require an AI to be perfect?

Some of the problems with AI stem from the nature of how the systems are built. An AI is based on a "model" – a collection of neurons that represent little pieces of knowledge that together form the algorithm, the functioning of the AI. The knowledge changes over time as the AI "learns" using feedback from external events, much like a child learns to ride a bicycle. While a computer that programs itself and improves over time seems great, it also causes concerns. In many cases we have no idea how successful AI systems function – they are a "black box" that operates in mysterious ways.

Other problems come from the data that the AI is trained with. It can be useful, sometimes essential, to provide an AI with large amounts of initial data to learn from, such as, thousands of X-rays that may contain cancer cells, millions of articles or social media posts on a particular subject, or hours of video footage of traffic situations. And if that data is tainted or skewed in a certain way, the AI will mimic that bias. For example, in 2016 Microsoft released Tay, one of the earliest general purpose chatbots. It was trained using millions of posts on public fora such as Quora and Reddit – places where anyone can post anything, and which are notorious for unfiltered content. After some initial success, Tay quickly started using rude and racist language and had to be taken offline[3].

---

[3] https://en.wikipedia.org/wiki/Tay_(bot)

The most sophisticated AI systems need massive amounts of data and computing power. Or to put it another way, those with access to big data, AI expertise and deep pockets to fund the computational needs will win in AI. That concentrates the power of this exceptional technology in the hand of a few companies (which are currently in the United States and in China). The increase in use of AI in our society risks exacerbating that imbalance of power, which will have geopolitical implications.

AI is already being used extensively in the military. Some nations are already using autonomous drones that kill humans, and the usage will inevitably increase. An AI arms race might very well be the logical result.

Of course, AI is just a tool, and like any tool, it can be used for good or bad purposes. However, as philosopher Nick Bostrom posited with his Vulnerable World Hypothesis[4], AI may be a technology that allows any group that is sufficiently advanced to destroy the world, regardless of the action of other players. Similar technologies are nuclear weapons, biotechnology and, perhaps, nanotechnology. As such AI requires special attention.

We are currently using what is known as artificial narrow intelligence (ANI), a form of AI that is still quite specific to certain tasks. In the future we may develop artificial general intelligence (AGI) that reaches a level of flexibility and general purpose equal to humans

> **What's a Model?**
>
> For an AI to function, it needs to have a model of the world. That can be done is several ways, but a popular method is a Neural Network. Taking inspiration from the human brain, a neural network consists of small entities - artificial neurons - that fire signals at each other and react to those signals. Together these neurons represent the algorithm that determines what the system does. And this algorithm - the way the neurons react- can change over time - either because they are manipulated by a developer or because the AI "learns" from feedback on its actions. This network of neurons can have many layers and is typically very large - billions of neurons in some models. Model interpretability - understanding what layers in the network represent and why they function the way they do, is then a difficult question.
>
> 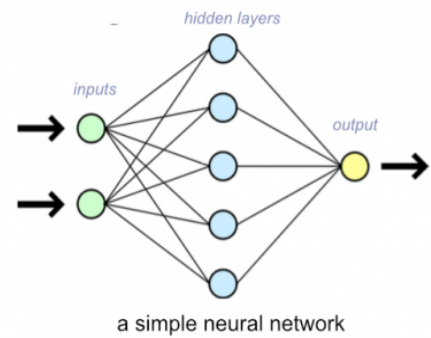
>
> a simple neural network

– to be able to react to any situation and solve any task. Some even speculate about artificial super intelligence (ASI) that far surpasses human cognitive abilities. Most researchers agree that AGI is decades, if not centuries, away. Of course, even if it is the former, we still need to take an

---

interest. But it may not help to worry about specifics, since we have so little idea of what will happen in the meantime and what the world will look like at that point in the future. The best we can do is build good ethical frameworks for now, and in the future.

## B. An overview of the issues

Leaving the geopolitical considerations aside, let us attempt to list the concerns systematically.

1. **Incompetence**

   This means the AI simply failing in its job. The consequences can vary from unintentional death (a car crash) to an unjust rejection of a loan or job application.

2. **Loss of privacy**

   AI offers the temptation to abuse someone's personal data, for instance to build a profile of them to target advertisements more effectively. While this may be an issue related to data, the nature of AI exacerbates the effects and also makes the situation more complex. For example, in the United States, a company called Clearview AI has collected publicly available photos of users from Facebook, LinkedIn and other platforms into a large database coupled with AI-driven facial recognition algorithms. The company is selling the system to law enforcement agencies. In Europe, the General Data Protection Regulation (GDPR) would clearly regulate this form of use of personal data, but the rules are much less clear when it comes to building a profile of a user to target advertisements, for instance.

   > **The connection between AI and Data**
   >
   > We have seen some of the issues with AI actually pertain to the data it was trained with. The connection between AI and data is important. Data shapes the algorithm any characteristics of the data will manifest in the AI. Issues in AI systems are similar to those in data governance, including data privacy, consent, purpose and intent, proportionality and protection. When we establish ethical principles, use design tools and methodologies or monitor and assess operational systems, we need to look at the data and the (AI) algorithm together.

3. **Discrimination**

   When AI is not carefully designed, it can discriminate against certain groups. An example is the algorithms the company Palantir developed for the Los Angeles Police Department[5]. The purpose of the AI was to support "predictive policing" – predict where crimes will happen so the police can proactively patrol. The AI was criticized for perpetuating systemic racism as, based on historical data, it flagged certain neighborhoods as higher risk. More recently, Chinese company Dahua was criticized for its facial recognition software that can specifically detect people from certain ethnic backgrounds[6].

---

[5] https://www.technologyreview.com/2020/07/17/1005396/predictive-policing-algorithms-racist-dismantled-machine-learning-bias-criminal-justice/

4. **Bias**

    The AI will only be as good as the data it is trained with. If the data contains bias (and much data does), then the AI will manifest that bias, too. If the AI is trained with data from a particular group only, it will start making the wrong assumptions. For example, a recent study showed that datasets from India drawn from online sources gives a skewed picture, since half of the population – particularly women and rural residents – do not have access to the internet.

5. **Erosion of Society**

    AI is used by many platforms that people get their daily news from. In the past, we used to choose from a range of TV channels and a range of newspapers. Many of us heard or read the exact same news stories. With online news feeds, both on websites and social media platforms, the news is now highly personalized for us. We risk losing a shared sense of reality, a basic solidarity. What makes things worse is that this hyper-personalization is not just for our benefit (as many platforms claim) – the main goal is to keep us engaged with the platform. As Tufekci observed in 2018[7], the AI that powers YouTube found out that people are drawn to content that is similar to what they started consuming but taken to a further extreme. In order to keep us glued, the AI suggests evermore extreme content. For example, politically conservative videos lead to extreme right-wing videos or conspiracy theories, or an interest in jogging leads to content about ultra-marathons. The AI has no evil intent per se - it is simply trying to achieve its objective of keeping you engaged.

6. **Lack of transparency**

    The idea of a "black box" making decisions without any explanation, without offering insight in the process, has a couple of disadvantages: it may fail to gain the trust of its users and it may fail to meet regulatory standards such as the ability to audit.

7. **Deception**

    AI has become very good at creating fake content. From text to photos, audio and video. The name "Deep Fake" refers to content that is fake at such a level of complexity that our mind rules out the possibility that it is fake. We all know that a photo can be altered. But a photo of a person that is artificially created by an AI is indistinguishable from a real photo. Audio and video can be faked to sophisticated levels, and we are not used to taking that possibility in to

---

[6]   https://en.wikipedia.org/wiki/Dahua_Technology

[7]   Tufekci, Z. (2018). YouTube, the great Radicalizer. Retrieved March 7, 2021.

https://www.nytimes.com/2018/03/10/opinion/sunday/youtube-politics-radical.html

account. In 2020, hundreds of millions of dollars were stolen by convincing company employees to make transfers using the deep-faked voice of their CEO.

8. **Unintended consequences**

Sometimes an AI finds ways to achieve its given goals in ways that are completely different from what its creators had in mind. Imagine a home robot is tasked to keep the house clean realizes that the family cat tends to make the house messy and decides killing the cat is a good way towards its given objective.

9. **Manipulation**

The 2016 scandal involving Cambridge Analytica is the most infamous example where people's data was crawled from Facebook and analytics were then provided to target these people with manipulative content for political purposes. While it may not have been AI per se, it is based on similar data and it is easy to see how AI would make this more effective.

10. **Lethal Autonomous Weapons (LAW)**

As the use of AI is becoming ubiquitous, increased use of AI-powered weapons may be inevitable. What is debated as an ethical issue is the use of LAW — AI-driven weapons that fully autonomously take actions that intentionally kill humans. Technologically, the possibility clearly already exists, whether the final decision ("pulling the trigger") is left to a human is a policy choice. The UN has deliberated the issue[8] and a number of groups and some nations have argued for a ban of LAW. The UN is the prime forum for such a discussion[9].

> **Should some AI be kept secret?**
>
> Is certain AI technology too dangerous to make publicly available? Should it be restricted, similar to what is done in the field of IT security, synthetic biology and nuclear physics? Some researchers argue for that.
>
> The opposite argument is that if powerful technology is in the hands of a few only that creates imbalance and increases potential for abuse. OpenAI was set up with the vision to develop research and make it available to everyone, thereby leaving the playing field. However, it ended up developing advanced algorithms that require such large computational resources that only a handful of powerful players would benefit from it, so they decided not to release it publicly.
>
> In general, academia plays a large role in the advances in AI and there is much openness in that community - new techniques are typically shared within months or even weeks.

---

[8] UNGA. "First Committee Weighs Potential Risks of New Technologies as Members Exchange Views on How to Control Lethal Autonomous Weapons, Cyberattacks". https://www.un.org/press/en/2018/gadis3611.doc.htm

[9] Gill (2020) "The Role of the United Nations in Addressing Emerging Technologies in the Area of Lethal Autonomous Weapons Systems" https://www.un.org/en/un-chronicle/role-united-nations-addressing-emerging-technologies-area-lethal-autonomous-weapons

11. **Malicious use of AI**

    Just as AI can be used in many different fields, it is unfortunately also helpful in perpetrating digital crimes. AI-supported malware and hacking are already a reality. There is a 2020 report by EUROPOL, UNICRI and Trend Micro on this topic[10].

12. **Loss of Autonomy**

    Respect for personal freedom and autonomy is ingrained into basic human rights in many societies. Delegating decisions to an AI, especially an AI that is not transparent and not contestable, may leave people feeling helpless, subjected to the decision power of a machine.

13. **Exclusion**

    The best AI techniques requires a large amount resources: data, computational power and human AI experts. There is a risk that AI will end up in the hands of a few players, and most will lose out on its benefits. This would have both societal and geopolitical implications.

Another ethical issue is the **Humane treatment of AI**. Given that AI is an algorithm, it may seem odd to worry about its treatment. But as AI becomes increasingly sophisticated and humanoid robots are developed that look like us and behave like us, the discussion will continue. A number of recent science fiction movies explore this theme.

---

[10]   Trend Micro, Europol, UNICRI. 2020. "Malicious Uses and Abuses of Artificial Intelligence"
http://www.unicri.it/node/3278

# 2. Defining Ethical AI

## C. What is Ethics?

We want our AI to be "good", or at least to avoid all the negative characteristics we have identified. We need a comprehensive set of ethical principles that will guide the development and use of AI. Unfortunately, there is no single truth for ethics.

Ethics are the standards of right and wrong, acceptable and not acceptable, in a certain community. It includes an individual's rights, obligations, virtues, the benefits to society, and a sense of fairness. Ethics will clearly differ for various cultures or groups. What is acceptable in one country may not be in another. They also change over time. What was commonly accepted 100 years ago is not OK today.

Even for the basic theory of ethics, different approaches exist. Classical philosopher Aristotle proposed Virtue Ethics ("What kind of person should I be?"). Modern philosopher Immanuel Kant wrote about Deontological Ethics ("How should I behave?"). Another approach is Consequentialist Ethics ("What should be my goal?"). Those are only a few of the many other theories of ethics.

Some basic ethical principles go to the core of the questions we are trying to solve about AI. For instance, Aristotle writes in the Nicomachean Ethics[11] about moral responsibility and argues that "the action must have its origin in the agent". These ideas become concretely relevant once we grant AI systems autonomy (that is, *agency*). Who is responsible for an accident caused by an autonomous car? The owner? The dealer that sold the car? The manufacturer of the AI? The supplier of the data that trained the AI?

We can see our ethical values as being composed of different layers, going from global to group to individual: *Core ethical values* based on inalienable human rights (human dignity, autonomy); *constitutional values* (rule of law, equity, privacy); *group specific* values (beliefs or cultural norms); and *individual ethical values* (personal convictions).

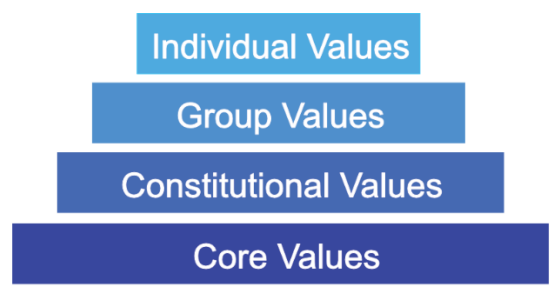

---

[11]   Aristotle. 350 BCE. "Nicomachean Ethics". Available as free e-book online. See also https://en.wikipedia.org/wiki/Nicomachean_Ethics"

We can observe the differences in ethics in various countries and cultures in the emphasis they place in policies and regulations. A recent study shows the differences between individualist and collectivistic cultures when it comes to moral dilemmas[12]. Much of the discussion (and regulation) about data privacy concerns come from countries with highly individualistic cultures. Conversely, Confucianism teaches that the society is more important than the individual and this is engrained in the mind of people steeped in this philosophy. At the same time, Buddhism regards the "self" as an illusion, so one can wonder how relevant the data privacy of the individual then is. That is not to say there is no concept of a right to privacy or other rights of the individual in these cultures, but the emphasis is different.

A 2019 study of the global landscape of AI ethics guidelines[13] shows that the United States, European Union and United Kingdom are the most prolific producers of ethics guidelines. However, there are ethics guidelines being produced in South Korea and Japan as well. In its 2017 AI Plan[14], China states the goal to become the world leader in AI by 2030… but also "emerge as the driving force in defining ethical norms and standards for AI". Several Chinese institutions have since developed sets of principles.

Another cultural difference is the balance between protection of human rights and the progress of business. Let us not forget there is a lot of money at stake in AI. With its General Data Protection Regulation (GDPR), the EU has implemented some of the most stringent controls on personal data, but some complain that it stifles business opportunities.

## D. What basis do we have?

Agreeing on a set of ethical values is harder if the group of stakeholders is larger and more diverse. For the UN, the challenge is hardest as we need to serve all people. However, we do have a basis to work from: The UN Charter, the Universal Declaration of Human Rights (UDHR) and a number of international treaties give us a basis for AI principles. In fact, some principles may not need to be reinvented, just applied in the context of AI.

---

[12] Awad et al. 2018. "The Moral Machine Experiment". https://doi.org/10.1038/s41586-018-0637-6.

[13] Jobin, Anna, Marcello Ienca, and Effy Vayena. 2019. "Artificial Intelligence: The Global Landscape of Ethics Guidelines." https://arxiv.org/abs/1906.11668.

[14] China State Council. 2017. "New Generation Artificial Intelligence Development Plan" (新一代人工智能发展规划). http://www.gov.cn/zhengce/content/2017-07-20/content_5211996.htm

Ethical values and ideas from all cultures may have useful concepts to contribute to the healthy principles for AI. The IEEE Standards Association investigated, in its report "Ethically Aligned Design"[15], how applying Buddhist, Ubuntu, and Shinto-inspired ethics could improve responsible AI. A recent publication from MIT Technology Review explored what Buddhism may offer to the ethics of AI.[16] Some ideas from that paper are that "any ethical use of AI must strive to decrease pain and suffering" and the "Do no harm principle: the burden of proof would be with those seeking to show that a particular application of AI does not cause harm". Another study highlights ways in which issues surrounding AI in India differ from Western countries and may call for different approaches to achieve fairness[17].

---

[15] IEEE. 2019. "Ethically Aligned Design: A Vision for Prioritizing Human Well-being with Autonomous and Intelligent Systems, First Edition" . https://standards.ieee.org/content/ieee-standards/en/industry-connections/ec/autonomous-systems.html

[16] Soraj Hongladarom / MIT technology Review. 2021. "What Buddhism can do for AI ethics". https://www.technologyreview.com/2021/01/06/1015779/what-buddhism-can-do-ai-ethics/

[17] Sambasivan et al. 2021. "Re-imagining Algorithmic Fairness in India and Beyond". https://research.google/pubs/pub50002/

# E. Frameworks for Ethical AI

Many nations, groups or organizations have published well thought-out sets of AI principles that large teams of experts have worked on and each offers its own perspective. The sets of principles are entitled with a "brand-name" that gives an indication of the particular perspective the group has taken. A bewildering number of these go around and they are not always well defined and sometimes confused: Trustworthy AI; Responsible AI; Explainable AI; Interpretable; Transparent; Human-centered; Safe AI or Mindful AI; and more.

The brand names do not matter nearly as much as the principles that are contained in the frameworks; and we see the same principles appear in many of them. The box on the right makes an attempt at clarifying some of the terms used to describe the frameworks, although for most, no single definition exists. Below, we will dig deeper into a few significant ones.

The European Commission's High-Level Expert Group on Artificial Intelligence (AI HLEG) published one of the best know frameworks in its report "Guidelines for Trustworthy AI"[18]. The term **Trustworthy** here conveys a general sense of trust that humans may have in the AI because of AI that can be trusted by humans to be "lawful, ethical and robust", where lawful means

> **What's in a name?**
>
> A number of terms are used to describe frameworks of ethical principles for AI. These terms reflect the general mindset, the spirit of the principles.
>
> **Ethical AI** = ensures compliance with ethical norms.
>
> **Trustworthy AI** = can be trusted by humans as acting in a" lawful, ethical and robust" way (AI HLEG).
>
> **Explainable** (or explicable) **AI**, also written as XAI = allows its functioning to be explained to stakeholders in non-technical terms. Synonyms or closely related terms are: Interpretable, Comprehensible and Understandable AI.
>
> **Interpretable AI** = on the same spectrum as Explainability but adds the ability for stakeholders not only to see but also study the decision-making process of the AI.
>
> **Meaningful AI** = used in the Villani report (France) to describe a system that is explainable and environmentally friendly, does not increase exclusion or inequality.
>
> **Transparent AI** = provides some level of accessibility to the data or algorithm.
>
> **Responsible AI** = takes into account societal values, moral and ethical considerations.
>
> **Human-centered AI** = ensures that human values are central to the way in which AI systems are developed, deployed, used and monitored.
>
> **Beneficial AI** = does not only avoid risks but contributes positively to society. The notion of beneficial AI came from one of the Asilomar Principles.

---

[18] EC High-Level Expert Group On Artificial Intelligence. 2019. "Guidelines for Trustworthy AI". https://ec.europa.eu/digital-single-market/en/high-level-expert-group-artificial-intelligence

comply with the laws in the jurisdiction the AI operates or touches upon, ethical meaning following a set of ethical principles and robust meaning the AI system is programmed well so as not to malfunction and cause unintentional harm.

The Montreal Declaration[19] is aimed at creating **Responsible AI**. It contains a sense of societal responsibility, including the principles of *Wellbeing*, *Solidarity* and *Democratic participation*. It also mentions *Respect for autonomy* and *Prudence* (meaning to be cautious when developing AI, anticipate any possible negative effects). *Accountability* and *Transparency* are other key elements.

Several others, including the UK Royal Society, use the term **Explainable AI**. The focus here is on the ability to explain algorithmic decisions (and sometimes the data driving those decisions is included) to end users and other stakeholders in non-technical terms. It is the opposite of a "Black Box". The AI HLEG describes it as: "auditable and comprehensible and intelligible by human beings at varying levels of comprehension and expertise".

There are two types: *Model explainability* is describing how the AI works on the data as a whole; however, the model may often be too complex for humans to understand. *Sample explainability* is explaining a specific sample — how the AI came to that particular decision.

Explainability is not everything - as the Royal Society states: "Transparency and explainability of AI methods may […] be only the first step in creating trustworthy systems…". But explainability is contained as a principle in many frameworks and there are good arguments for it:

- *Validity*: Stakeholders need to be reassured of the validity and accuracy of the decisions.
- *Auditability:* Auditors require tools to check models for regulatory compliance.
- *Maintainability*: Data scientists need insight in their models to maintain and improve them.
- *Transparency*: Providing users and other stakeholders insight in the functioning of the AI is useful to earn their trust.

Some have mused whether Explainability might in the future become mandatory (like GDPR) in some countries or regions.

Explainable AI has an interesting potential: Often AI (specifically, Machine Learning) is used for complex environments that humans do not fully understand so we don't have a model. Weather patterns or stock price fluctuations are examples. If an AI can learn to predict those complex systems and is built as an Explainable AI, we might learn from the AI to understand the real-world phenomena better. This is sometime called Third-Wave AI or AI 3.0.

---

[19] see: https://www.montrealdeclaration-responsibleai.com/

One question that needs to be answered is: "Explainable to whom?" The flip side to explainability is the protection of intellectual property and strategic advantage. Commercial entities may not want to go fully explainable if the capabilities of the AI constitute their competitive advantage. And military users will likely not be keen to explain exactly how the technology in their AI-based weapons functions.

Organizations that formulate their ethical principles should start by asking themselves what their main objective is, the general spirit within which they want to work. They are free to use any of the terms mentioned here or make up their own description. And many do; you will find "Fair AI", "Mindful AI", "Accurate AI"… Remember, what truly matters are the principles.

## F. Principles

A large number of frameworks for ethical AI have been developed, each with a number of principles. We can find similar ideas in individual principles through all frameworks. Even though the names and the definitions may differ, the basic idea is the same.

The list of principles below is a collected from several frameworks and guidelines including those of the AI HLEG, IEEE, OECD, Villani report[20], ACM, the Japanese Society for AI, the Beijing AI principles[21], the Nation Science and Technology Council (US)[22], The Montreal Declaration and the Asilomar principles, organized in six categories:

---

[20] Villani et al. 2018. "For a meaningful artificial intelligence". https://www.aiforhumanity.fr/pdfs/MissionVillani_Report_ENG-VF.pdf.

[21] Beijing Academy of Artificial Intelligence. (2019). Beijing AI principles. https://www.baai.ac.cn/blog/beijing-ai-principles.

[22] Holdren / National Science and Technology Council. 2017. "Preparing for the Future of Artificial Intelligence" https://doi.org/10.1007/s00146-016-0685-0.

**Beneficence** ("do good")
- Wellbeing (societal and environmental)
- Solidarity
- Driving inclusive growth
- Sustainable development

**Non-maleficence** ("do no harm")
- Privacy
- Security and Safety
- Integrity

**Accountability**
- Transparency
- Lawfulness
- Equip AI with an 'ethical black box'
- Auditability

**Justice** ("be fair")
- Fairness
- Democracy
- Diversity, non-discrimination
- Ensure a genderless, unbiased AI.
- Equity
- Respect for Human Rights
- Access and redress
- Data Agency

**Competence** ("function well")
- Technical robustness and safety

**Governance** ("control the AI")
- Human agency and oversight

A 2020 report in Minds and Machines[23] has a more detailed overview of how each of these principles are represented in each of the frameworks and guidelines. An interesting observation is that while all principles are represented in some of the guidelines, none of the principles are represented in all guidelines. However, there seems to be most agreement on the principles of *accountability*, *privacy* and *fairness*, which are found in 80 percent of the guidelines.

How these principles from the various frameworks address the ethical concerns that we had identified earlier is shown in the following chart.

---

[23] Hagendorff. 2020. "The Ethics of AI Ethics: An Evaluation of Guidelines". https://doi.org/10.1007/s11023-020-09517-8.

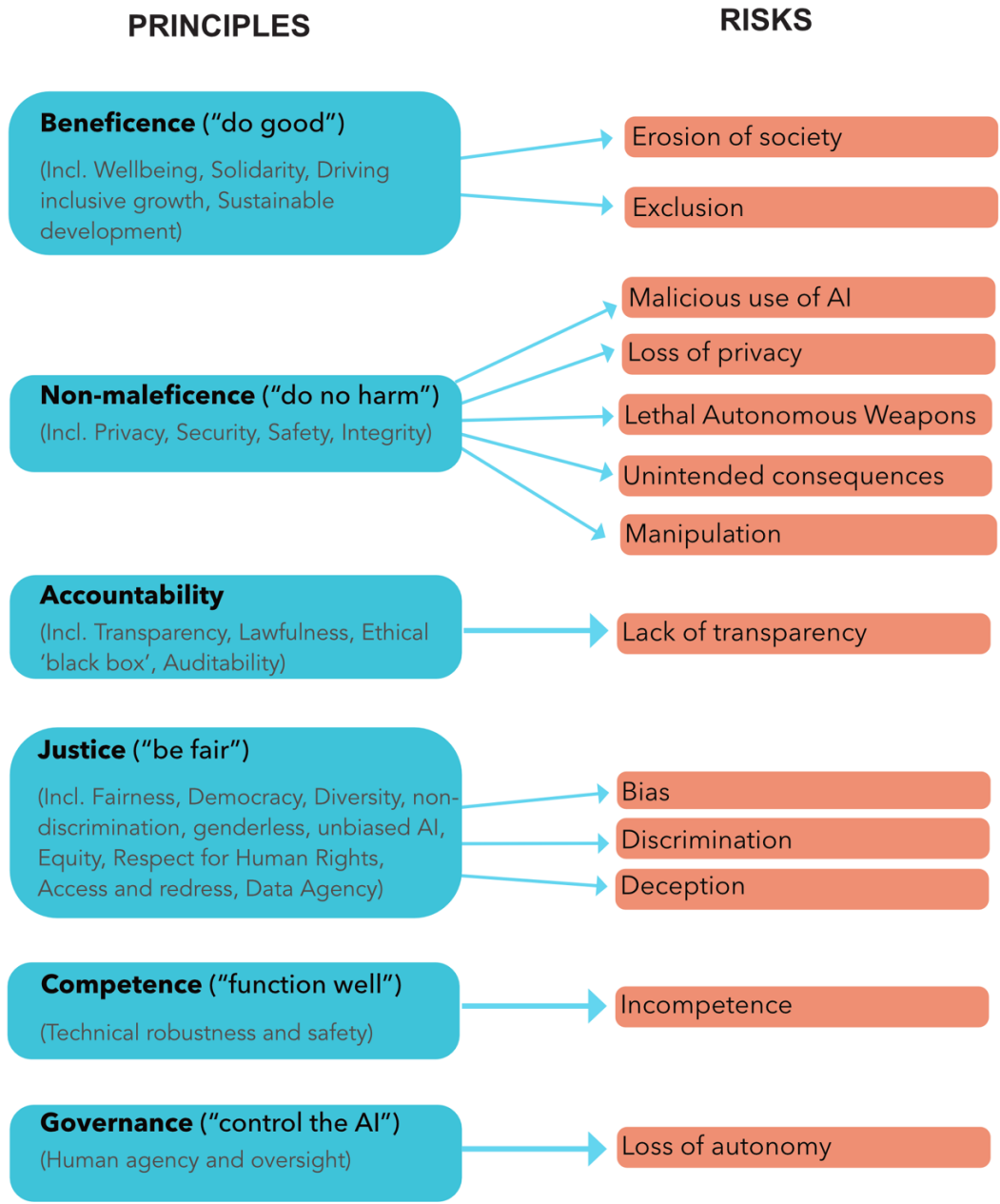

As we can see, the principles listed under *Non-maleficence* and *Justice* contribute to addressing a range of ethical concerns, while *Competence*, *Accountability* and *Governance* address single, but important concerns.

## G. Governance of Ethical AI policies

So far most ethical principles are used within a limited context — within a particular business or government agency. General regulation or legislation regarding AI is difficult as traditional mechanisms can be insufficient for a number of reasons[24]:

- Machines are not "people" (unless we define them as legal person at some point, just like businesses are[25]), and existing legal systems operate by assigning and allocating legal rights and responsibilities to "persons".
- The global nature of the internet and the opaque nature of many AIs means that risky AI development may easily escape (or intentionally evade) detection by regulators.
- AI is often made up of components created by different people at different places (in different jurisdictions) and times, without conscious coordination (e.g., on public code-sharing platforms such as Github).

The difficulty in monitoring or enforcing compliance with policies on Ethical AI is a concern that should be kept in mind when developing principles.

---

[24] Ideas taken from a presentation by Matt Sherer at Asilomar Conference. See: https://theengineeringofconsciousexperience.com/public-risk-management-for-ai-the-path-forward-matt-scherer/

[25] In 2017, the European Parliament suggested this, but it has not happened yet.

# 3. Implementing Ethical AI

## H. Operationalizing AI principles

Even if it may seem that our use of AI in the UN is still rather limited, and risks can perhaps easily be contained (probably not the case), it is clear in any case that our use of AI will become widespread and more complex in the coming years and that the ethical implications are not so easily foreseen.

Developing a set of ethical principles for AI is the first step. But they will not be realized if there is no support for those developing, purchasing, integrating, and using AI systems. Stakeholders need to be involved in the process of defining and operationalizing the Principles – the team that does this needs to be multi-disciplinary and represent stakeholder communities. It should also bring in different cultural perspectives (for which the UN, with its multi-national body of staff, is well positioned). The members of the team will need awareness and understanding of the issues, development methodologies, standardized assessment methodologies, and technical tools. The Principles could be provided as a voluntary guideline, a code of conduct or a policy. Sometimes new policies may not be needed if a reinterpretation of existing policies in the context of AI will suffice.

Aligning the development of AI systems with Principles can benefit from both technical and non-technical methods. On the technical side there are architectures, methodology, explanation methods, and technical tests. Non-technical methods include education, standardization, stakeholder involvement in design, and diversity in teams.

## I. Proposed Framework

For the UN to move toward ethical use of AI in the systems it develops and uses, requires a number of steps.

**Principles**: While there are many existing policies and frameworks to take inspiration from, the UN will need to articulate its own set of principles. These principles should be aligned with the approach we take to data, including, importantly, the CEB-approved Principles on Personal Data Protection and Privacy[26] and Data Governance policies and guidelines.

**Assessment methods**: Standard methods should be defined to assess an AI system's alignment with our principles. Several methods exist: The AI HLEG report includes a practical assessment list[27], the Z-Inspection process[28] is a well thought out method, based on the AI HLEG principles and the Canadian government has created an "Algorithmic Impact Assessment (AIA)" — an online questionnaire with 60 questions related to your business process, data and system designed decisions[29].

---

[26] See: https://unsceb.org/principles-personal-data-protection-and-privacy-listing

Assessments can be done both at the start of a project (example: a risk-assessment that will help decide whether the project should go ahead, and which risks need to be mitigated), before an AI is deployed or periodically on an AI that is in operation — to ensure they continue to meet the guidelines.

**Architectural Standards**: We need minimum standards that specify the technical requirements AI systems, whether developed in-house or acquired as product or service, must comply with. Certain principles such as *robustness* or *safety and security* would lend themselves particularly to close governance from an architectural perspective. The architectural standards would take a role in procurement as well. The WEF "AI Procurement in a box" toolkit[30] includes ways to integrate assessments into a procurement process.

**Development Tools and Methodologies:** These are meant to enable developers, integrators or users of AI systems to implement the principles. For instance, methodologies such as Ethics-by-Design and Value-Sensitive Design (VSD) can help to create AI in a way that leads to more accountability, responsibility and transparency. A recent paper proposes a modified form of VSD to use the Sustainable Development Goals (SDGs) as a base for ethical principles for AI that not only prevent harm but aim to do good[31]. In terms of tools, several vendors of AI services and platforms have realized the demand for ethically aligned development of AI and are offering specific tools. Many open-source tools also exist, such as LIME[32], a popular algorithm for AI explainability.

**Awareness and education**: This should include different levels. Anyone interacting with AI systems should be provided with information in non-technical language that explains the principles and our approach. Developers, data processors and certain users of the AI systems will need more detailed documentation on our principles including technical specifications as well general education in Ethics and training in specific tools or methodologies. Assessing the implications of an AI system can be far from trivial and specific education on this will be needed.

---

[27] See Chapter III of the report: https://ec.europa.eu/digital-single-market/en/high-level-expert-group-artificial-intelligence

[28] Zicari et al. 2021. "Z-Inspection: A Process to Assess Ethical AI". https://doi.org/10.1109/TTS.2021.3066209. See also: http://z-inspection.org/

[29] See:https://www.canada.ca/en/government/system/digital-government/digital-government-innovations/responsible-use-ai/algorithmic-impact-assessment.html

[30] See: https://www.weforum.org/reports/ai-procurement-in-a-box

[31] Umbrello and van de Poel. 2020. Mapping value sensitive design onto AI for social good principles. https://doi.org/10.1007/s43681-021-00038-3

[32] Available at https://homes.cs.washington.edu/~marcotcr/blog/lime/. See also an explanation at https://www.oreilly.com/content/introduction-to-local-interpretable-model-agnostic-explanations-lime/.

**Governance**: Our Ethical AI principles should result in a policy in the UN entity where it is implemented. Enforcement of this policy could happen as part of a project approval process, an architecture review board or it could be linked to data governance. The technically complex nature of AI systems will often require a balance between oversight (by auditors, senior management or other bodies) and the engineers developing or installing the AI.

## J. Effectiveness

Can we be sure that these efforts by developers and users of AI systems to implement the principles do indeed lead to AI that is safe, trustworthy, fair or to achieving our desired goals? Many companies have been criticized give lip-service to AI ethics ("ethics-washing") while little actually changes under the hood. But these companies have to balance their adherence to ethics with commercial interests. The UN, fortunately, does not have this tension.

Still, despite all well-intended efforts, the dividend may not be paid unless we are careful. What is key is not only to institute a policy, standards, methodologies and tools but importantly to foster a mindset among staff that both understands and is committed to the spirit of the principles. In that respect, communication and education must be an important element in our path forward.

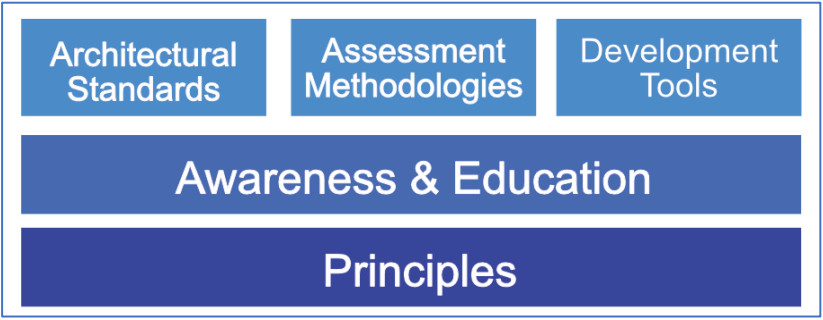

*Governance Model*

# Conclusion

AI technology has progressed much in the past few years and will continue to do so. Combined with the fact that our life is increasingly digitized, AI is becoming a powerful tool that will soon be ubiquitous in all types of devices and services we use in everyday life.

The UN is already using AI and the scope and complexity of its use will likely increase significantly in the coming years. There are, however, numerous ethical concerns related to the development and use of AI and we should plan carefully how we want to embrace it. To mitigate negative effects of an AI, we should establish a framework of principles for ethical AI.

Those principles need to align to the ethical values of the operators, owners, users and other stakeholders. In the case of the UN, we already have several frameworks to base our principles on, for instance the UN Charter and the Universal Declaration of Human Rights. Policies and frameworks have been developed by governments, regional organizations and research groups, all of which can help inform and inspire our set of principles.

AI is closely connected to data — the information that AI systems are fed with shape the algorithms. Therefore, ethical principles for AI need to closely align with those used for data, including the UN Principles on Personal Data Protection and Privacy.

Implementing our ethical principles, that is, ensuring that the AI we develop and use in the future will meet our ethical standards, requires a policy, standards, methodologies, and tools. But most importantly it requires awareness and education. The implementation of ethical AI principles all only succeed if their spirit is understood and adopted by all that interact with AI systems.